\documentclass[prd,showpacs,amsmath,amssymb,12pt]{revtex4}
\usepackage{bm}
\usepackage{slashed}
\usepackage{graphicx}
\usepackage{epsfig}
\usepackage[colorlinks, citecolor=blue]{hyperref}

\begin{document}

\title{Study the Forward-Backward Asymmetry of Top Quark Production in Warped Extra Dimension Model with an Extension of Strong Interaction}

\author{Cheng Li and Cai-Dian L\"{u}}
\email{lucd@ihep.ac.cn}
\affiliation{Institute of High Energy Physics and Theoretical Physics Center for Science Facilities, Chinese Academy of Sciences, Beijing 100049, People¡¯s Republic of China}
\author{Xiang-Dong Gao}
\affiliation{Institute of Physics, Academia Sinica, Nangang, Taipei 11529, Taiwan}

\begin{abstract}
  The large forward-backward asymmetry of top quark pair production measured by hadron colliders shed light on new physics signals beyond the Standard Model.
  In the warped extra dimension model with an additional $SU(3)$ group in strong sector, we compare the cross section and forward-backward asymmetry of top quark pair production with recent data obtained by CDF and D0. 
  Our numerical analysis shows that the parameter $c_q \geq 0.5$, $c_t \sim -0.6$, $\tan\phi \geq 20$ and the first excitation of axial gluon with a mass about $5 \thicksim 6 \mathrm{TeV}$ can accommodate this large anomaly without violating other experimental constraints.
  We show that a large ratio of strong couplings $g^D/g^S$ sets a strong limit on the parameter space of this model.
\end{abstract}



\maketitle

\section{INTRODUCTION}

Top quark physics is one of the most attractive areas of current elementary particle physics, since it is the heaviest fermion in the Standard Model (SM). 
And it is also assumed having a large coupling to new physics particles in many extensions of SM.
Hadron colliders provide a crucial tool for uncovering the mystery of electroweak symmetry breaking within top quark, although the most accurate measurement, such as total production cross section at the Tevatron \cite{Aaltonen:2010bs, Abazov:2009si} is in good agreement with theoretical predictions in the standard model \cite{Moch:2008qy, Cacciari:2008zb, Kidonakis:2010dk, Ahrens:2011px}. 
Recently, the forward-backward asymmetry has been measured by the CDF and D0 collaborations \cite{Aaltonen:2011kc, Abazov:2011rq}:
\begin{eqnarray}
CDF: \ 20.1 \pm 6.7\% ,\nonumber\\
D0: \ 19.6 \pm 6.5\% .
\end{eqnarray}

Theoretically, this forward-backward asymmetry of top quark pair production from $q \bar{q}$ annihilation at hadron colliders, vanishes at leading-order QCD calculation in inclusive $t \bar{t}$ production.
Due to the interference of gluon exchange tree diagram with box diagram and interference between initial and final state gluon radiation, the next-to-leading-order QCD contribution to forward-backward asymmetry is about 6\% \cite{Kuhn:1998jr, Kuhn:1998kw, Almeida:2008ug, Kuhn:2011ri}, while the electroweak correction can only contribution roughly 0.2 \% \cite{Hollik:2011ps}.
This indicates that the current experimental measurement of this asymmetry is more than 2 standard deviation from the SM expectations \cite{Bowen:2005ap}, and even $3.4 \sigma$ effect has been claimed for $M_{t \bar{t}} > 450 \mathrm{GeV}$:
\begin{eqnarray}
A_{FB}^{t \bar{t}} = 0.475 \pm 0.014 .
\label{eq2}
\end{eqnarray}
The ATLAS and CMS collaborations did not discover any deviation from the SM predictions within the experimental errors \cite{ATLAS:2012an, Chatrchyan:2011hk}, because initial states are almost charge symmetric due to gluon fusion at the LHC.

Right after this experimental indication, a series of new physics models have been proposed to solve this problem, such as an additional color-octet gauge boson \cite{Ferrario:2008wm, Gabrielli:2011jf, Gabrielli:2011zw}, a new t-channel physics contribution \cite{Cheung:2009ch, Jung:2009jz}, a supersymmetric singlet \cite{delaPuente:2011iu}, and so on \cite{Cao:2010zb}.
In Ref. \cite{Frampton:2009rk} a flavor-nonuniversal chiral color model has been discussed, where  a parity violating light axigluon $M_A \sim 1.5 \mathrm{GeV}$ is preferred to give large forward-backward asymmetry.

The warped extra dimension model which is proposed by Randall and Sundrum  \cite{Randall:1999ee, Randall:1999vf}, is one of the most compelling candidates for extensions of SM, not only can
explain the gauge hierarchy problem but also the fermion flavor puzzle in an elegant geometric manner.  
In this model, the Kaluza-Klein gluons can generate forward-backward asymmetry at tree level.
Unfortunately, it is noted in ref. \cite{Djouadi:2009nb} that the exchange of Kaluza-Klein gluons at tree level is difficult to generate a remarkable forward-backward asymmetry, without violating other phenomenological constraints.

In this paper, we will extend the warped extra dimension model with an extension of strong interaction to include the Kaluza-Klein axial-gluon.  
Similar to Ref. \cite{Frampton:2009rk}, we find that the axial-gluon contribution in the above models can also accommodate the experimental measured forward-backward asymmetry, without violating other phenomenological constraints.
In section \textrm{II}, we present the model explicitly and point out the importance of axigluon that induces forward-backward asymmetry at the Tevatron. 
We show the formulae of cross section and charge asymmetry in section \textrm{III}. And the numerical results can be found in section \textrm{IV}, which can explain the data obtained by CDF and D0. 
Finally, we give a brief summary in section \textrm{V}.

\section{THE MODEL}

In the Randall-Sundrum (RS) model, two 3-branes located at $0$ (Planck scale) and $k \pi R$ (TeV scale) on the fifth dimension, where the bulk space is a slice of anti-de Sitter ($\mathrm{AdS_5}$) space with curvature $k$ and radius $R$. 
If we let the SM particles propagate in the bulk except Higgs particle, which is assumed to localize around TeV brane, we can avoid the dangerous higher dimensional operators, which can induce sizable flavor-changing-neutral-currents \cite{Grossman:1999ra, Gherghetta:2000qt}. 
The zero modes of fermion can be localized either on the infrared (IR) brane or the ultraviolet (UV) brane, which only depend on the mass parameters of different flavors \cite{Huber:2003tu}.
Since top quark is much heavier than the others, we naturally assume its zero mode is localized on the IR-brane with the mass parameter $c_t < \frac{1}{2}$, while the other light quarks are localized on the UV-brane with the mass parameter $c_q > \frac{1}{2}$.

In literature \cite{Bauer:2011ah}, an additional $SU(3)$ group is added in strong sector, which enlarges the strong interaction group to $SU(3)_D \times SU(3)_S$, where subscripts $D$ and $S$ indicate couplings to $SU(2)_L$ quark doublets and singlets respectively. 
In this model, the so-called RS flavor problem \cite{Csaki:2008zd}, which refers to a fine-tuning of Kaluza-Klein mass to meet the CP-violating observable $\epsilon_K$ in $K - \bar{K}$ oscillation, can be solved beautifully without violate the Randall-Sundrum-Glashow-Iliopoulos-Maiani (RS-GIM) mechanism \cite{Agashe:2004ay}.
The five-dimensional color-octet gauge gluons $G^D_M$ and $G^S_M$ couple to $SU(2)_L$ quark doublets and singlets, respectively, where the spacetime subscript $M$ runs from 0 to 5. 
The interaction Lagrangian of gluons and quarks is
\cite{Bauer:2011ah}
\begin{eqnarray}
\mathcal{L}_{int} = \sum_{r=1}^{family=3} g^D \bar{Q}_{rL} \Gamma^M G^D_M Q_{rL} + g^S \bar{U}_{rR} \Gamma^M G^S_M U_{rR} + g^S \bar{D}_{rR} \Gamma^M G^S_M D_{rR},
\end{eqnarray}
where the five-dimension Dirac matrices $\Gamma^M = e^M_N \gamma^N$,  $e^M_N$ are funfbeins and $\gamma^N = (\gamma^{\mu},i \gamma^5)$.

A tiny modification of minimal bulk gauge group $SU(3)_C \times SU(2)_L \times U(1)_Y$ is needed, where custodial symmetry is usually proposed in electroweak sector \cite{Agashe:2003zs} to meet the electroweak phenomenology (such as the oblique parameters $S$, $T$ and the $Z^0 b \bar{b}$ coupling) \cite{Davoudiasl:1999tf, Csaki:2002gy, Burdman:2002gr} and flavor phenomenology (such as the $K$, $D$ and $B$ physics) \cite{Blanke:2008zb, Blanke:2008yr, Albrecht:2009xr}. 
As a consequence, the Kaluza-Klein (KK) mass $M_{KK} > 2.4 \mathrm{TeV}$ is required \cite{Casagrande:2008hr, Bauer:2009cf}.

We should rotate the gauge group to recover five-dimensional standard model gluons $G_M = \cos\phi G^D_M + \sin\phi G^S_M$. 
As a byproduct, the axigluons $A_M = -\sin\phi G^D_M + \cos\phi G^S_M$ emerge, where $\tan\phi \equiv g^D/g^S$. 
The five-dimensional strong coupling is $g_s^{(5)} = g^D \cos\phi = g^S \sin\phi$ and four-dimensional strong coupling is $g_s = g_s^{(5)}/\sqrt{\pi R}$. 
$\pi R$ is the length of the orbifold, and $k \pi R \sim 37$ for stabilizing the gauge hierarchy from Planck scale. 
So we can express the action of axigluons and fermions as:
\begin{eqnarray}
S^A_{int} = \int d^4x dy \sqrt{-g} \sum_{i=1}^{flavor=6} (-\tan\phi \bar{\Psi}_{iL}(x,y) \Gamma^M \Psi_{iL}(x,y) + \nonumber \\ 
\cot\phi \bar{\Psi}_{iR}(x,y) \Gamma^M \Psi_{iR}(x,y))A_M(x,y).
\end{eqnarray}

After KK decomposition and fixing the gauge $A_M(x,y)=(A_{\mu}(x,y),0)$, we obtain
\begin{eqnarray}
S^A_{int} = \sum_{i=1}^{flavor=6} \sum_{n=1}^{\infty} \int d^4x g_s (-\tan\phi \alpha^i_{nL} \bar{\Psi}_{iL}^{(0)}(x) \gamma^{\mu} \Psi_{iL}^{(0)}(x) + \nonumber \\
\cot\phi \alpha^i_{nR} \bar{\Psi}_{iR}^{(0)}(x) \gamma^{\mu} \Psi_{iR}^{(0)}(x)) A^{(n)}_{\mu}(x),
\end{eqnarray}
where $\Psi_{iL,R}^{(0)}(x)$ are fermion zero modes, and $A^{(n)}_{\mu}(x)$ are KK excitations of axigluons which have no zero modes naturally. 
The entanglement functions of gauge boson and fermion profiles are \cite{Gherghetta:2000qt}
\begin{eqnarray}
\alpha^{i,(n)}_{L,R} = \int_0^{\pi R} dy \frac{2(1 \mp c_{iL,R})}{e^{(1 \mp 2c_{iL,R})k \pi R}-1} e^{2(1 \mp c_{iL,R})ky} \frac{k}{N^{(n)}} (J_1(\frac{M_A^{(n)}}{k} e^{ky}) + b_1(M_A^{(n)}) Y_1(\frac{M_A^{(n)}}{k} e^{ky})),
\label{eq4}
\end{eqnarray}
where $b_1(M_A^{(n)}) \sim 0$ and $N^{(n)} \sim 1/\sqrt{\pi^2 R M_A^{(n)} e^{\pi kR}}$.

We take account only the first excitation of axigluon for simplicity, whose mass is $M_A \sim 2.4 M_{KK}$\cite{Bauer:2011ah}. 
For a saving of parameters and simplicity, we assume $c_{iL} = -c_{iR} = c_i$, thus $\alpha^{i,(n)}_L = \alpha^{i,(n)}_R$, and denote $c_q$ and $\alpha_q$ for light quarks while $c_t$ and $\alpha_t$ for top quark, we can obtain:
\begin{eqnarray}
\mathcal{L}^A_{int} = \sum_{i=1}^{light \  quarks} g_s \bar{f}_i \gamma^{\mu} (f_L P_L + f_R P_R) f_i A_{\mu} + g_s \bar{t} \gamma^{\mu} (g_L P_L + g_R P_R) t A_{\mu},
\end{eqnarray}
where $f_L = -\tan\phi \alpha_q$ and $f_R = \cot\phi \alpha_q$, while $g_L = -\tan\phi \alpha_t$ and $g_R = \cot\phi \alpha_t$, and projection operators are $P_{L,R} = \frac{1 \mp \gamma^5}{2}$.
The emergence of massive axigluon is curial to the top quark production asymmetry, since it violates parity symmetry in strong sector. 
In another word, it induces forward-backward asymmetry within CP conservation.  
We will show that in the next section.

The detail of symmetry breaking to QCD gauge group is model dependent, but as a minimal possibility we assume it occurs in two steps on IR-brane with proper boundary conditions.
$SU(3)_D \times SU(3)_S \times SU(2)_L \rightarrow SU(3)_C \times SU(2)_L$ through the zero mode of scalar particle $\Phi$, which transforms as $(\textbf{3}, \bar{\textbf{3}}, \textbf{1})$, and the vacuum expectation value of $\langle \Phi^{(0)} \rangle$ is much larger than $\mathcal{O}(1) \mathrm{TeV}$.
Followed by $SU(2)_L \times U(1)_Y \rightarrow U(1)_{EM}$ due to the usual Higgs doublet $\phi^{(0)}$, which transforms as $\textbf{2}_{1/2}$ and vacuum expectation value is $v \sim 246 \mathrm{GeV}$.
And we ignore the potential custodial symmetry breaking.

Although the colored Higgs particle $\Phi^{(0)}$ could induce the dangerous proton decay, we assume its mass is heavy enough to suppress it.
Since $\Phi^{(0)}$ induces a scalar current that is parity conservation, there is no additional contribution to forward-backward asymmetry of top quark pair production.
And the contribution to Peskin-Takeuchi parameter can be tamed by adding the custodial symmetry on weak sector \cite{Agashe:2003zs}.
The Kaluza-Klein modes of Higgs particles $\Phi^{(n)}$ and $\phi^{(n)}$ are localized near the IR-brane, but their effects are quite tiny due to the large masses which cannot be detected by the current high energy experiments.

\section{CROSS SECTION AND FORWARD-BACKWARD ASYMMETRY}

\begin{figure}[tb]
  \begin{center}
  \begin{tabular}{cc}
    \hspace{-3cm}
    \includegraphics[scale=0.67]{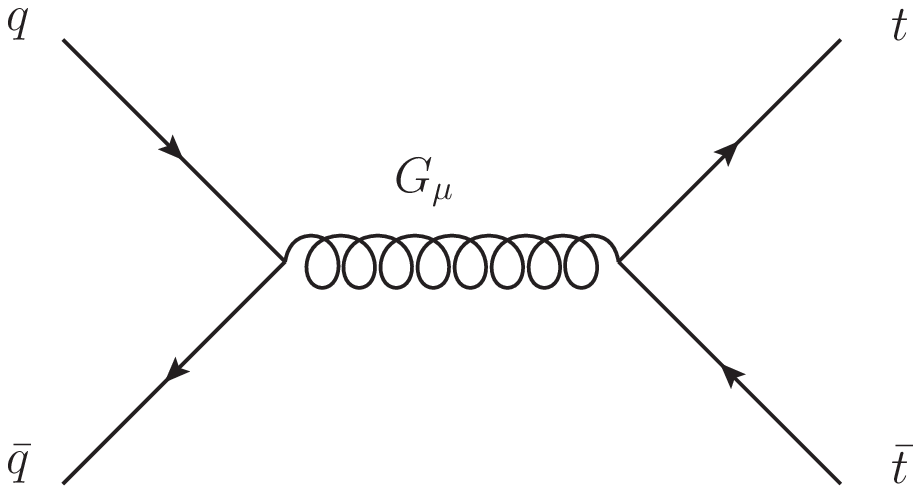} & \hspace{-7cm} \includegraphics[scale=0.67]{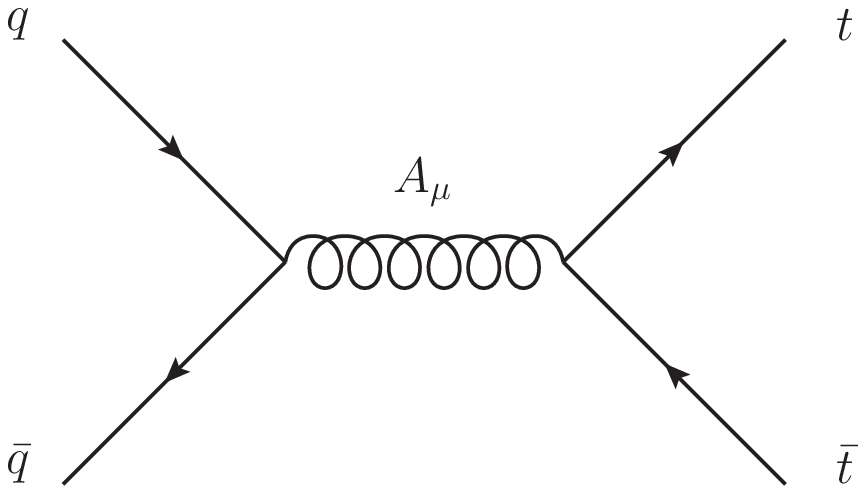} \\
    \vspace{-15cm}
  \end{tabular}
  \caption{The left Feynman diagram is QCD leading order contribution for $t \bar{t}$ inclusive production, and the right one is axigluon exchange at tree level.}
  \label{1} 
  \protect
  \end{center}
\end{figure}

In the Tevatron $p \bar{p}$ collider, the top quark pair can be produced though the leading Feynman diagram of the SM as the fist one in Fig. \ref{1}. 
In the current model, there is one additional diagram exchanging axigluon shown as the second one in Fig. \ref{1}. 
Since the cross section measured by the experiments agree well with the SM calculations, the new physics contribution should be much smaller than the standard model one.
The interference between the QCD Born diagram and s-channel axigluon exchange will induce the forward-backward asymmetry, since the axigluon breaks parity conservation.

In partonic center-of-mass frame the differential cross section is \cite{Cao:2010zb}:
\begin{eqnarray}
\frac{d \hat{\sigma}}{d \cos\theta} = \frac{d \hat{\sigma}_{SM}}{d \cos\theta} + \frac{d \hat{\sigma}_{INT}}{d \cos\theta} + \frac{d \hat{\sigma}_{NP}}{d \cos\theta},
\end{eqnarray}
where $\theta$ is top quark polar angle in the center-of-mass frame. 
The QCD leading order contribution which is charge symmetric is
\begin{eqnarray}
\frac{d \hat{\sigma}_{SM}}{d \cos\theta} = \frac{\pi \beta \alpha_s^2}{9 \hat{s}} (2- \beta^2 + \beta^2 \cos^2\theta).
\end{eqnarray}
The interference between gluon and axigluon contributes
\begin{eqnarray}
\frac{d \hat{\sigma}_{INT}}{d \cos\theta} = \frac{\pi \beta \alpha_s^2}{18 \hat{s}} \frac{\hat{s}}{\hat{s} - M_A^2} [(g_L + g_R)(f_L + f_R)(2- \beta^2) + \nonumber \\ 
2(g_L - g_R)(f_L - f_R) \beta \cos\theta + (g_L + g_R)(f_L + f_R) \beta^2 \cos^2\theta].
\end{eqnarray}
Finally the third term induced by axigluons only is
\begin{eqnarray}
\frac{d \hat{\sigma}_{NP}}{d \cos\theta} = \frac{\pi \beta \alpha_s^2}{36 \hat{s}} \frac{\hat{s}^2}{(\hat{s} - M_A^2)^2} [(g_L^2 + g_R^2)(f_L^2 + f_R^2)(1+ \frac{2g_L g_R}{g_L^2 + g_R^2}(1-
\beta^2)) + \nonumber \\
2(g_L^2 - g_R^2)(f_L^2 - f_R^2) \beta \cos\theta + (g_L^2 + g_R^2)(f_L^2 + f_R^2) \beta^2 \cos^2\theta].
\end{eqnarray}
The partonic center-of-mass energy is $\sqrt{\hat{s}} = \sqrt{x_1 x_2 s}$, where $x_1, x_2$ are longitudinal fractions of proton or anti-proton momentum carried by parton 1 and 2, respectively.
$\sqrt{s}$ is the total center-of-mass  energy of the hadron collider. 
And $\beta = \sqrt{1- \frac{4m_t^2}{\hat{s}}}$ is top quark velocity in the $t \bar{t}$ center-of-mass frame.
Here we assume that the mass of axigluon is larger than colliding energy $\sqrt{s}$ of the Tevatron by considering other constraints.

One should convolute these cross sections with parton distribution functions  to obtain the right cross section in the laboratory frame with a proper boost and cut. 
The forward-backward asymmetry of the top quark pair productions is  defined as
\begin{eqnarray}
A_{FB}^{p \bar{p}} = \frac{\sigma_F - \sigma_B}{\sigma_F + \sigma_B} = \frac{\sigma_a}{\sigma_s},
\end{eqnarray}
where the symmetric and anti-symmetric cross sections at partonic level are defined as:
\begin{eqnarray}
\hat{\sigma}_{s,a}(\hat{s}) = \int_0^1 d \cos\theta \frac{d \hat{\sigma}}{d \cos\theta} \pm \int_{-1}^0 d \cos\theta \frac{d \hat{\sigma}}{d \cos\theta}.
\end{eqnarray}

\begin{figure}[tb]
  \begin{center}
  \begin{tabular}{cccc}
    \hspace{-2.8cm}
    \includegraphics[scale=0.51]{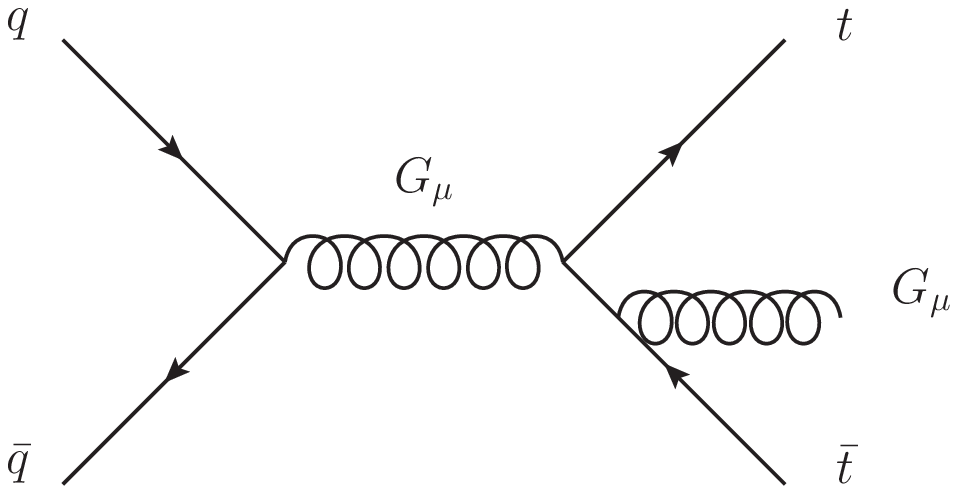} & \hspace{-7.4cm} \includegraphics[scale=0.51]{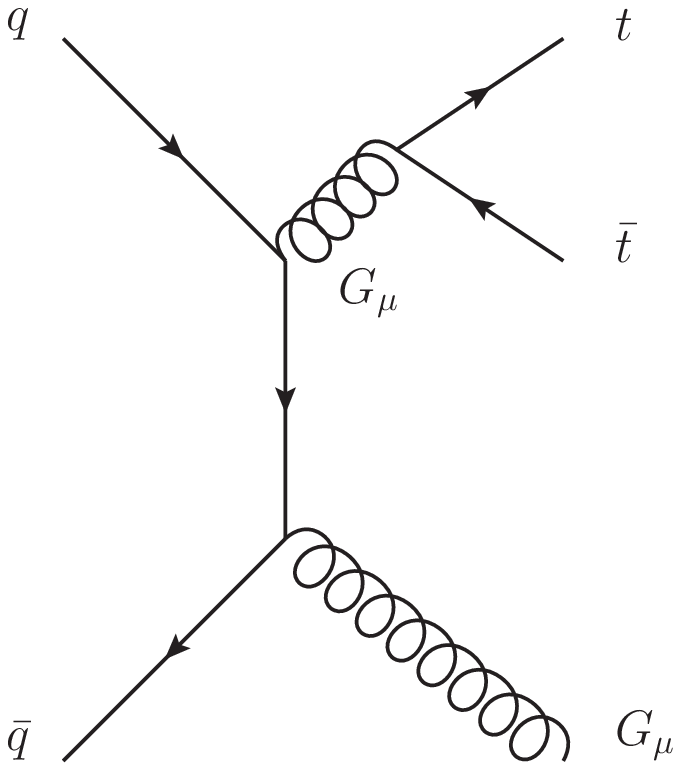} & \hspace{-7cm} \includegraphics[scale=0.51]{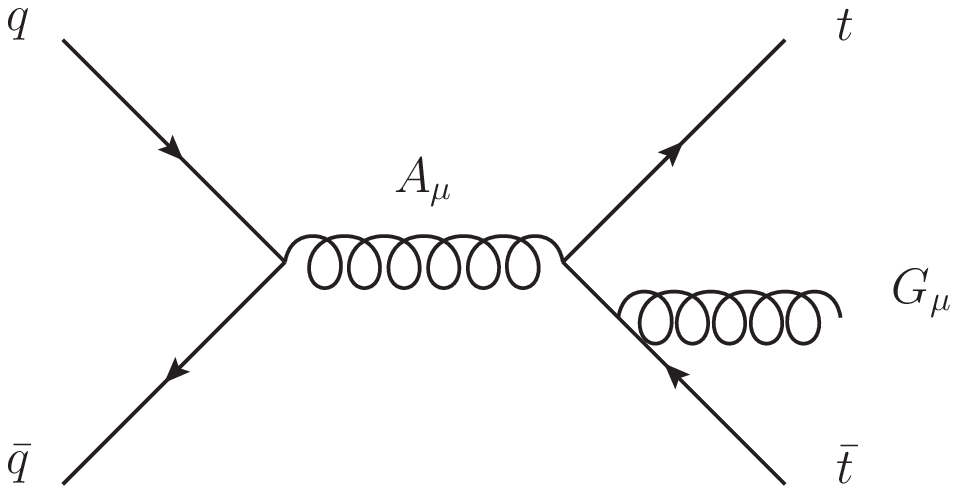} & \hspace{-7.4cm} \includegraphics[scale=0.51]{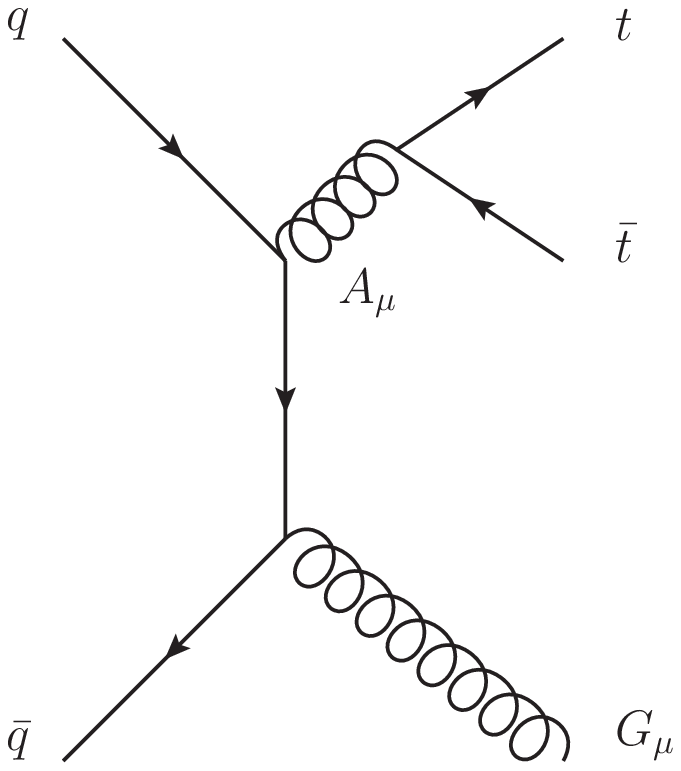}
    \vspace{-10cm}
  \end{tabular}
  \caption{The  leading order  QCD s-channel and t-channel  Feynman diagrams in $t \bar{t} +j$ exclusive production and the s-channel and t-channel axigluon exchange diagrams at leading order.}
  \label{2} 
  \protect
  \end{center}
\end{figure}

From Eq. (\ref{eq2}), we notice that a proper cut on the invariant mass of $t \bar{t}$ can enhance the signal/background efficiency in experimental measurements.
Theoretically, we need consider the $t \bar{t}$ plus jet production. 
The typical Feynman diagrams are shown in Fig. \ref{2}. 
Practically, the invariant mass of top-antitop pair $M_{t \bar{t}}$ can be expressed by jet energy $E_j$ at leading-order:
\begin{eqnarray}
M_{t \bar{t}}^2 = \hat{s} (1- \frac{2E_j}{\sqrt{\hat{s}}}).
\end{eqnarray}

\section{NUMERICAL RESULTS}

In numerical analysis, we use the parton distribution functions of CTEQ5 \cite{Pumplin:2002vw} to obtain a hadronic level results.
For the extra dimension model  with custodial symmetry $SU(3)_C \times SU(2)_L \times SU(2)_R \times U(1)_{B-L}$, the fit of oblique $S$, $T$ parameters indicates $M_{KK} \sim \mathcal{O}(1) \mathrm{TeV}$ \cite{Agashe:2003zs}. 
Since the strong gauge sector of our model is $SU(3)_D \times SU(3)_S$ which is not custodial symmetry on Higgs sector, so there is no more contribution to Peskin-Takeuchi parameters \cite{Peskin:1990zt, Peskin:1991sw}.
The flavor-changing-neutral-currents in $B$ physics is suppressed by the expansion factor $\nu^2/M_{KK}^2$, and current data show that $k \pi R \simeq 37$, $c_t \in [-0.3, -1]$, $c_q \in [0.5, 0.8]$ and $M_{KK} > 1.2 \mathrm{TeV}$ are allowed \cite{Casagrande:2008hr, Bauer:2009cf, Albrecht:2009xr}.
We calculate forward-backward asymmetry of top quark pair production within strong $SU(3)_D \times SU(3)_S$ model, but the model with custodial symmetry $SU(2)_L \times SU(2)_R \times U(1)_{B-L}$ helps us to choose a rough range of parameters in the beginning of calculation.

\begin{figure}[tb]
  \begin{center}
  \begin{tabular}{cc}
    \hspace{0cm}
    \includegraphics[scale=1.0]{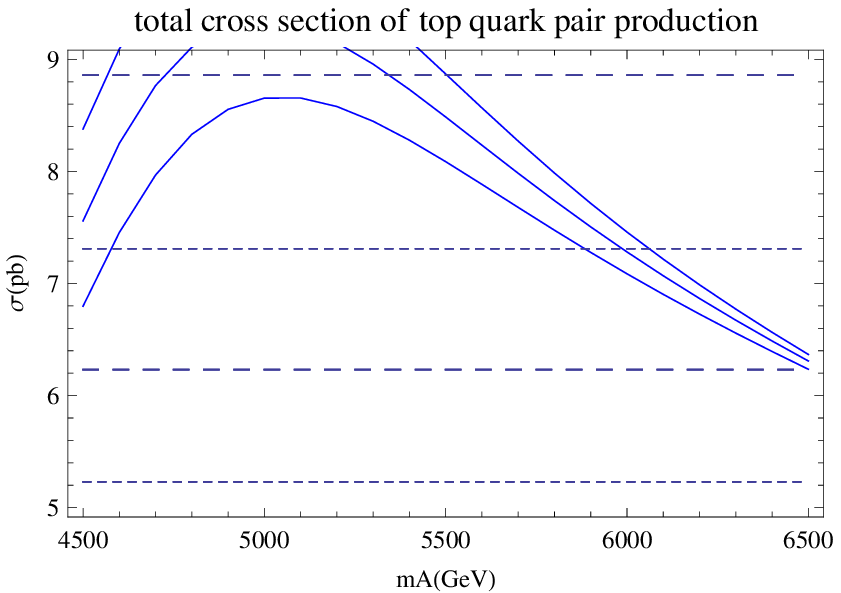} & \includegraphics[scale=1.07]{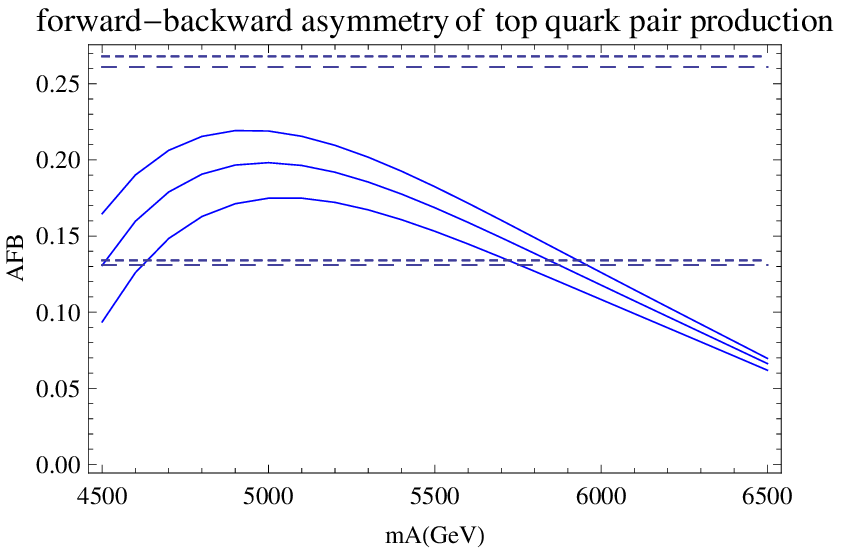}
    \vspace{0cm}
  \end{tabular}
  \caption{The total cross section (left) and the forward-backward asymmetry (right) of top quark pair production as a function of axial gluon mass. The three curves from bottom to top in the pictures are for $c_t = -0.60$, $c_t = -0.70$, $c_t = -0.80$, respectively. 
  Other parameters are chosen as $c_q = 0.51$, and $\tan\phi = 25$.
  The dashed lines are CDF $1 \sigma$ allowed bands and dotted ones are D0 data.}
  \label{3} 
  \protect
  \end{center}
\end{figure}

Firstly, the experimental measured total cross sections will give a severe constraint of the model parameters. 
We show the $t \bar{t}$ cross section as a function of the axial gluon mass $M_A$ in the left panel of Fig. \ref{3}, with  the mass parameters of light quarks $c_q=0.51$ and top quark $c_t=-0.60$, $-0.70$ and $-0.80$ (three lines from bottom to top).
The two dashed lines in the figure are the CDF experimental $1 \sigma$ band and dotted ones are the D0 $1 \sigma$ band.
At the right panel of Fig. \ref{3}, we also show the forward-backward asymmetry of $t \bar{t}$ production as a function of $M_A$ using the same model parameters.
Here we take $\tan\phi = 25$ to obtain a sizable asymmetries which will be explained below.
From Fig. \ref{3}, one can see that there is a maximum in the region of $5 \thicksim 6 \mathrm{TeV}$ for the mass of axigluon, preferred by the $t \bar{t}$ experiments, which indicate $M_{KK} > 2 \mathrm{TeV}$. 
The axial gluon mass dependence behavior of Fig. \ref{3} reflects the entanglement Bessel functions in Eq. (\ref{eq4}). 
Since the coupling of axigluon to top quark is much larger than that to the light quarks, the forward-backward asymmetry is enhanced only in the $t \bar{t}$ final states.

\begin{figure}[tb]
  \begin{center}
  \begin{tabular}{cc}
    \hspace{0cm}
    \includegraphics[scale=1.0]{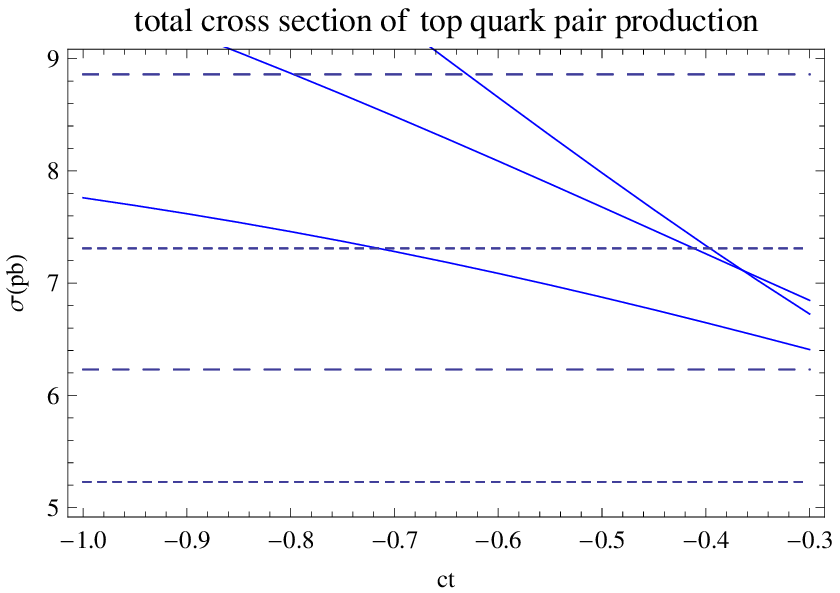} & \includegraphics[scale=1.07]{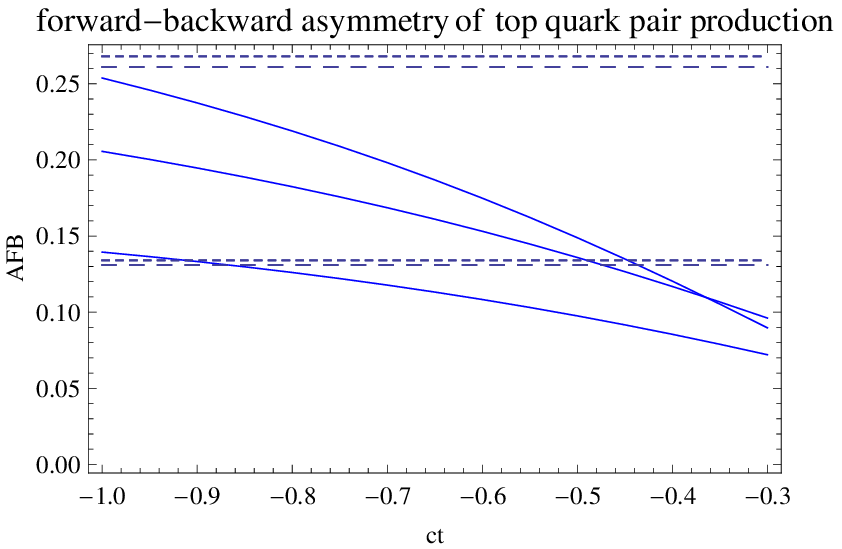}
    \vspace{0cm}
  \end{tabular}
  \caption{The left picture is total cross section of top quark pair production, and the right one is forward-backward asymmetry. 
  The parameters are chosen for $c_q=0.51$, $M_A=5 \mathrm{TeV}$, $5.5 \mathrm{TeV}$ and $6 \mathrm{TeV}$ and $\tan\phi=25$.
  The dashed lines are CDF bands and dotted ones are D0 data.}
  \label{4} 
  \protect
  \end{center}
\end{figure}

The smallness of light quark profiles around IR-brane is less important for the problem, so we consider top quark mass parameter only. 
The difference between light quarks and top quark coupling to the axigluon is the origin of the forward-backward asymmetry in top quark pair production. 
The hierarchy of fermion masses will increase the forward-backward asymmetry of top quark pair production.
When $c_t$ increases, in other words, the mass of top quark decreases, the entanglement function is getting smaller and smaller. 
Therefore we obtain decent curves of both cross section and forward-backward asymmetry within running $c_t$. 
This is explicitly shown in Fig. \ref{4} as a function of $c_t$ with $M_A = 5 \mathrm{TeV}$, $5.5 \mathrm{TeV}$ and $6 \mathrm{TeV}$, with $c_q = 0.51$ and $\tan\phi = 25$. 
The left picture is total cross section of top quark pair production, and the right one is forward-backward asymmetry. 
We use the dashed lines for CDF experimental bands and dotted ones for D0 data.

\begin{figure}[tb]
  \begin{center}
  \begin{tabular}{cc}
    \hspace{0cm}
    \includegraphics[scale=1.0]{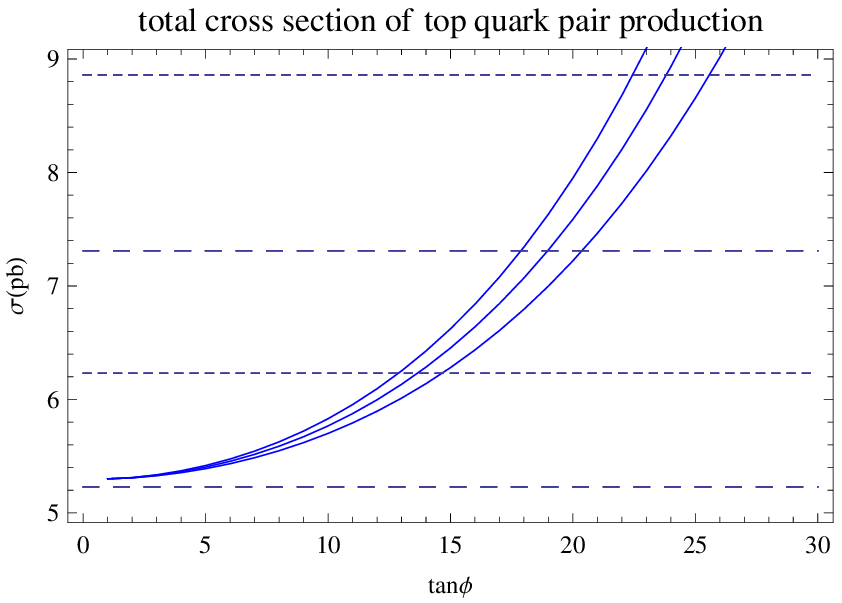} & \includegraphics[scale=1.07]{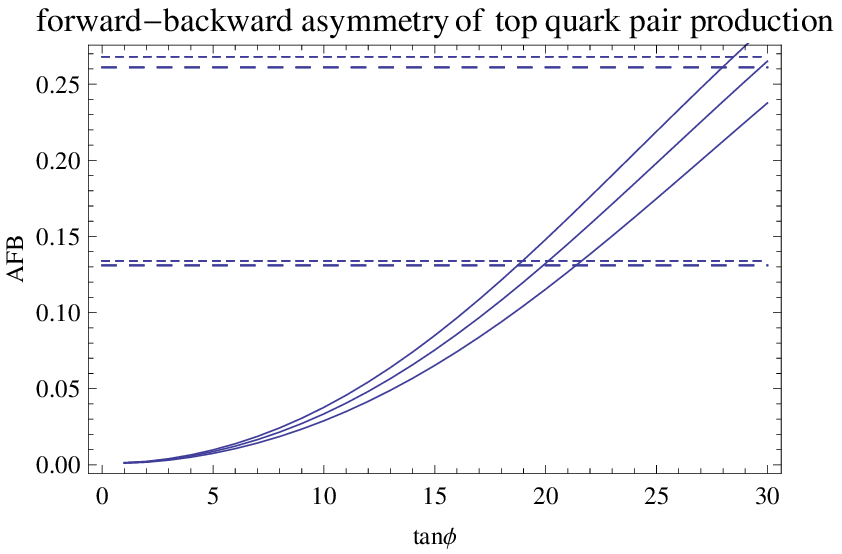}
    \vspace{0cm}
  \end{tabular}
  \caption{The left picture is total cross section of top quark pair production, and the right one is forward-backward asymmetry. 
  The parameters are chosen for $c_q=0.51$, $c_t=-0.60$, $-0.70$ and $-0.80$, and $M_A=5 \mathrm{TeV}$.
  The dashed lines are CDF bands and dotted ones are D0 data.}
  \label{5} 
  \protect
  \end{center}
\end{figure}

Next we show the cross section and forward-backward asymmetry as a function of $\tan\phi$ in Fig. \ref{5}, which is much important for the model building. 
Here we set $M_A=5 \mathrm{TeV}$, $c_q=0.51$ and $c_t=-0.60, -0.70, -0.80$. 
The left picture in Fig. \ref{5}, is the total cross section of top quark pair production, and the right one is forward-backward asymmetry. 
The dashed lines express CDF experimental bands and dotted ones express D0 data. 
From Fig. \ref{5}, one can see that $\tan\phi$ around $\mathcal{O}(1)$ is not preferred.
And $\tan\phi > 20$ is allowed for a sizable forward-backward asymmetry in $t \bar{t}$ production, which is not accordant with the assumption in \cite{Bauer:2011ah}. 
This means that $g^D \gg g^S$, so the gauge groups are hierarchy in the bulk. 
On the other hand, the computation is perturbative, if we set $g^D \sim 1$ and $g^S \ll 1$. 
In this case, the parity violation is very tiny even in the higher dimensions. 
We will have only one $SU(3)$ group in the strong sector at low energy. 
Additionally, since the forward-backward asymmetry is proportional to $\tan\phi + \frac{1}{\tan\phi}$ which is symmetric under transformation $\tan\phi \rightarrow 1/ \tan\phi$, $\tan\phi \leq \frac{1}{25}$ is also acceptable. 
However, the hierarchy of gauge group is unchanged.

\begin{figure}[tb]
  \begin{center}
  \includegraphics[scale=1.0]{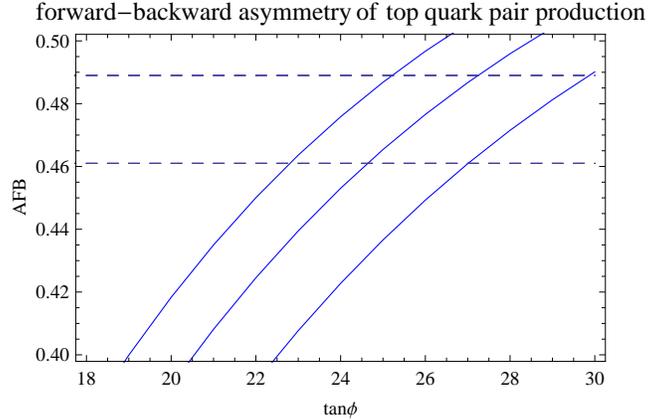}
  \caption{The forward-backward asymmetry of $t \bar{t} + jet$ final state. 
  The parameters are chosen for $c_q=0.51$, $c_t=-0.60$, $-0.70$ and $-0.80$ and $M_A = 5 \mathrm{TeV}$. 
  The horizontal band is obtained by CDF $1 \sigma$ data.}
  \label{6} 
  \protect
  \end{center}
\end{figure}

Finally, as stated in the introduction, more effective experimental data for the forward-backward asymmetry is presented from CDF collaboration for the $t \bar{t} + jet$ final states. 
Our results are shown in Fig. \ref{6}, using a $M_{t \bar{t}} > 450 \mathrm{GeV}$ cut to compare with the experiment data. 
Here we set $c_q=0.51$, $c_t=-0.60$, $-0.70$ and $-0.80$ and $M_A = 5 \mathrm{TeV}$. 
The horizontal band in the figure is obtained from the CDF $1 \sigma$ data. 
We are particular interested in the forward-backward asymmetry running with $\tan\phi$, which is consistent with the results shown in Fig. \ref{5}.
From Fig. \ref{6} we can see that with one additional hard jet, the constraint to new physics parameters are more stringent, since the asymmetry has been enhanced.
A careful study will provide more information from exclusive production than the inclusive one.

In Ref. \cite{Frampton:2009rk} a general flavor-nonuniversal chiral color model has been discussed, where a light axigluon $M_A \sim 1.5 \mathrm{GeV}$ and a moderate mixing angle $\tan\phi \sim 30^\circ$ are preferred.
However, since the other experimental constraints on warped extra dimension model existed \cite{Casagrande:2008hr,Bauer:2009cf,Albrecht:2009xr}, our mass of axigluon is much larger.
In words, a special parameter space is obtained for $c_q \geq 0.5$, $c_t \sim -0.6$, $M_A = 5 \thicksim 6 \mathrm{TeV}$ and $\tan\phi \geq 20$, which is consistent with other constraints.
Therefore the parameter space of this model is severe constrained, which can be tested easily by experiment in future. 
The axigluons is crucial to induce forward-backward asymmetry in $p \bar{p}$ collision since it violates parity, and the difference of coupling to top quark from light quarks boost it significantly. 
The contributions of higher Kaluza-Klein excitations are suppressed by $M_{KK}$, so we can ignore them at present.

\section{SUMMARY}

We compute total cross sections and forward-backward asymmetries of top quark pair production in Randall-Sundrum model with enlarging strong sector to $SU(3)_D \times SU(3)_S$. 
Utilyzing the recent CDF and D0 data of forward-backward asymmetry, a special parameter space is obtained for $c_q \geq 0.5$, $c_t \sim -0.6$, $M_A = 5 \thicksim 6 \mathrm{TeV}$ and $\tan\phi \geq 20$, which is also consistent with flavor experiments and electroweak precision tests.

\section{ACKNOWLEDGMENT}

This work is partially supported by National Science Foundation of China under the Grant No. 11228512, 11235005 and 11075168.


\begin{thebibliography}{99}

\bibitem{Aaltonen:2010bs}
  T.~Aaltonen {\it et al.}  [CDF Collaboration],
  Phys.\ Rev.\ D {\bf 82}, 052002 (2010)  [arXiv:1002.2919 [hep-ex]].  

\bibitem{Abazov:2009si}
  V.~M.~Abazov {\it et al.}  [D0 Collaboration],
  Phys.\ Lett.\ B {\bf 679}, 177 (2009)  [arXiv:0901.2137 [hep-ex]].  

\bibitem{Moch:2008qy}
  S.~Moch and P.~Uwer,
  Phys.\ Rev.\ D {\bf 78}, 034003 (2008)  [arXiv:0804.1476 [hep-ph]].  

\bibitem{Cacciari:2008zb}
  M.~Cacciari, S.~Frixione, M.~L.~Mangano, P.~Nason and G.~Ridolfi,
  JHEP {\bf 0809}, 127 (2008)  [arXiv:0804.2800 [hep-ph]].  

\bibitem{Kidonakis:2010dk}
  N.~Kidonakis,
  Phys.\ Rev.\ D {\bf 82}, 114030 (2010)  [arXiv:1009.4935 [hep-ph]].  

\bibitem{Ahrens:2011px}
  V.~Ahrens, A.~Ferroglia, M.~Neubert, B.~D.~Pecjak and L.~L.~Yang,
  Phys.\ Lett.\ B {\bf 703}, 135 (2011)  [arXiv:1105.5824 [hep-ph]].  

\bibitem{Aaltonen:2011kc}
  T.~Aaltonen {\it et al.}  [CDF Collaboration],
  Phys.\ Rev.\ D {\bf 83}, 112003 (2011)  [arXiv:1101.0034 [hep-ex]].  

\bibitem{Abazov:2011rq}
  V.~M.~Abazov {\it et al.}  [D0 Collaboration],
  Phys.\ Rev.\ D {\bf 84}, 112005 (2011)  [arXiv:1107.4995 [hep-ex]].  

\bibitem{Kuhn:1998jr}
  J.~H.~Kuhn and G.~Rodrigo,
  Phys.\ Rev.\ Lett.\  {\bf 81}, 49 (1998)  [hep-ph/9802268].  

\bibitem{Kuhn:1998kw}
  J.~H.~Kuhn and G.~Rodrigo,
  Phys.\ Rev.\ D {\bf 59}, 054017 (1999)  [hep-ph/9807420].  

\bibitem{Almeida:2008ug}
  L.~G.~Almeida, G.~F.~Sterman and W.~Vogelsang,
  Phys.\ Rev.\ D {\bf 78}, 014008 (2008)  [arXiv:0805.1885 [hep-ph]].  

\bibitem{Kuhn:2011ri}
  J.~H.~Kuhn and G.~Rodrigo,
  JHEP {\bf 1201}, 063 (2012)  [arXiv:1109.6830 [hep-ph]].  

\bibitem{Hollik:2011ps}
  W.~Hollik and D.~Pagani,
  Phys.\ Rev.\ D {\bf 84}, 093003 (2011)  [arXiv:1107.2606 [hep-ph]].  

\bibitem{Bowen:2005ap}
  M.~T.~Bowen, S.~D.~Ellis and D.~Rainwater,
  Phys.\ Rev.\ D {\bf 73}, 014008 (2006)  [hep-ph/0509267].  

\bibitem{ATLAS:2012an}
  G.~Aad {\it et al.}  [ATLAS Collaboration],
  Eur.\ Phys.\ J.\ C {\bf 72}, 2039 (2012)  [arXiv:1203.4211 [hep-ex]].  

\bibitem{Chatrchyan:2011hk}
  S.~Chatrchyan {\it et al.}  [CMS Collaboration],
  Phys.\ Lett.\ B {\bf 709}, 28 (2012)  [arXiv:1112.5100 [hep-ex]].  

\bibitem{Ferrario:2008wm}
  P.~Ferrario and G.~Rodrigo,
  Phys.\ Rev.\ D {\bf 78}, 094018 (2008)  [arXiv:0809.3354 [hep-ph]].  

\bibitem{Gabrielli:2011jf}
  E.~Gabrielli and M.~Raidal,
  Phys.\ Rev.\ D {\bf 84}, 054017 (2011)  [arXiv:1106.4553 [hep-ph]].  

\bibitem{Gabrielli:2011zw}
  E.~Gabrielli, M.~Raidal and A.~Racioppi,
  Phys.\ Rev.\ D {\bf 85}, 074021 (2012)  [arXiv:1112.5885 [hep-ph]].  

\bibitem{Cheung:2009ch}
  K.~Cheung, W.~-Y.~Keung and T.~-C.~Yuan,
  Phys.\ Lett.\ B {\bf 682}, 287 (2009)  [arXiv:0908.2589 [hep-ph]].  

\bibitem{Jung:2009jz}
  S.~Jung, H.~Murayama, A.~Pierce and J.~D.~Wells,
  Phys.\ Rev.\ D {\bf 81}, 015004 (2010)  [arXiv:0907.4112 [hep-ph]].  

\bibitem{delaPuente:2011iu}
  A.~de la Puente,
  JHEP {\bf 1202}, 016 (2012)  [arXiv:1111.4488 [hep-ph]].  

\bibitem{Cao:2010zb}
  Q.~-H.~Cao, D.~McKeen, J.~L.~Rosner, G.~Shaughnessy and C.~E.~M.~Wagner,
  Phys.\ Rev.\ D {\bf 81}, 114004 (2010)  [arXiv:1003.3461 [hep-ph]].  

\bibitem{Frampton:2009rk}
  P.~H.~Frampton, J.~Shu and K.~Wang,
  Phys.\ Lett.\ B {\bf 683}, 294 (2010)  [arXiv:0911.2955 [hep-ph]].  

\bibitem{Randall:1999ee}
  L.~Randall and R.~Sundrum,
  Phys.\ Rev.\ Lett.\  {\bf 83}, 3370 (1999)  [hep-ph/9905221].  

\bibitem{Randall:1999vf}
  L.~Randall and R.~Sundrum,
  Phys.\ Rev.\ Lett.\  {\bf 83}, 4690 (1999)  [hep-th/9906064].  

\bibitem{Djouadi:2009nb}
  A.~Djouadi, G.~Moreau, F.~Richard and R.~K.~Singh,
  Phys.\ Rev.\ D {\bf 82}, 071702 (2010)  [arXiv:0906.0604 [hep-ph]].  

\bibitem{Grossman:1999ra}
  Y.~Grossman and M.~Neubert,
  Phys.\ Lett.\ B {\bf 474}, 361 (2000)  [hep-ph/9912408].  

\bibitem{Gherghetta:2000qt}
  T.~Gherghetta and A.~Pomarol,
  Nucl.\ Phys.\ B {\bf 586}, 141 (2000)  [hep-ph/0003129].  

\bibitem{Huber:2003tu}
  S.~J.~Huber,
  Nucl.\ Phys.\ B {\bf 666}, 269 (2003)  [hep-ph/0303183].  

\bibitem{Bauer:2011ah}
  M.~Bauer, R.~Malm and M.~Neubert,
  Phys.\ Rev.\ Lett.\  {\bf 108}, 081603 (2012)  [arXiv:1110.0471 [hep-ph]].  

\bibitem{Csaki:2008zd}
  C.~Csaki, A.~Falkowski and A.~Weiler,
  JHEP {\bf 0809}, 008 (2008)  [arXiv:0804.1954 [hep-ph]].  

\bibitem{Agashe:2004ay}
  K.~Agashe, G.~Perez and A.~Soni,
  ``B-factory signals for a warped extra dimension,''  Phys.\ Rev.\ Lett.\  {\bf 93}, 201804 (2004)  [hep-ph/0406101].  

\bibitem{Agashe:2003zs}
  K.~Agashe, A.~Delgado, M.~J.~May and R.~Sundrum,
  JHEP {\bf 0308}, 050 (2003)  [hep-ph/0308036].  

\bibitem{Davoudiasl:1999tf}
  H.~Davoudiasl, J.~L.~Hewett and T.~G.~Rizzo,
  Phys.\ Lett.\ B {\bf 473}, 43 (2000)  [hep-ph/9911262].  

\bibitem{Csaki:2002gy}
  C.~Csaki, J.~Erlich and J.~Terning,
  Phys.\ Rev.\ D {\bf 66}, 064021 (2002)  [hep-ph/0203034].  

\bibitem{Burdman:2002gr}
  G.~Burdman,
  Phys.\ Rev.\ D {\bf 66}, 076003 (2002)  [hep-ph/0205329].  

\bibitem{Blanke:2008zb}
  M.~Blanke, A.~J.~Buras, B.~Duling, S.~Gori and A.~Weiler,
  JHEP {\bf 0903}, 001 (2009)  [arXiv:0809.1073 [hep-ph]].  

\bibitem{Blanke:2008yr}
  M.~Blanke, A.~J.~Buras, B.~Duling, K.~Gemmler and S.~Gori,
  JHEP {\bf 0903}, 108 (2009)  [arXiv:0812.3803 [hep-ph]].  

\bibitem{Albrecht:2009xr}
  M.~E.~Albrecht, M.~Blanke, A.~J.~Buras, B.~Duling and K.~Gemmler,
  JHEP {\bf 0909}, 064 (2009)  [arXiv:0903.2415 [hep-ph]].  

\bibitem{Casagrande:2008hr}
  S.~Casagrande, F.~Goertz, U.~Haisch, M.~Neubert and T.~Pfoh,
  JHEP {\bf 0810}, 094 (2008)  [arXiv:0807.4937 [hep-ph]].  

\bibitem{Bauer:2009cf}
  M.~Bauer, S.~Casagrande, U.~Haisch and M.~Neubert,
  JHEP {\bf 1009}, 017 (2010)  [arXiv:0912.1625 [hep-ph]].  

\bibitem{Pumplin:2002vw}
  J.~Pumplin, D.~R.~Stump, J.~Huston, H.~L.~Lai, P.~M.~Nadolsky and W.~K.~Tung,
  JHEP {\bf 0207}, 012 (2002)  [hep-ph/0201195].  

\bibitem{Peskin:1990zt}
  M.~E.~Peskin and T.~Takeuchi,
  Phys.\ Rev.\ Lett.\  {\bf 65}, 964 (1990).  

\bibitem{Peskin:1991sw}
  M.~E.~Peskin and T.~Takeuchi,
  Phys.\ Rev.\ D {\bf 46}, 381 (1992).  

\end{thebibliography}
\end{document}